\documentclass{appolb}
\usepackage{amsmath}
\usepackage{amsfonts,amsbsy}
\usepackage{amssymb}
\usepackage{graphicx}

\catcode`\@=11

\newcount\@tempcntc
\def\@citex[#1]#2{\if@filesw\immediate\write\@auxout{\string\citation{#2}}\fi
  \@tempcnta\z@\@tempcntb\m@ne\def\@citea{}\@cite{%
        \@for\@citeb:=#2\do%
    {\@ifundefined{b@\@citeb}%
        {\@citeo\@tempcntb\m@ne\@citea%
                \def\@citea{,\penalty\@m\ }{\bf ?}\@warning%
                {Citation `\@citeb' on page \thepage \space undefined}}%
        {\setbox\z@\hbox{\global\@tempcntc0\csname b@\@citeb\endcsname\relax}
     \ifnum\@tempcntc=\z@ \@citeo\@tempcntb\m@ne%
       \@citea\def\@citea{,\penalty\@m}%
       \hbox{\csname b@\@citeb\endcsname}%
     \else%
      \advance\@tempcntb\@ne%
      \ifnum\@tempcntb=\@tempcntc%
      \else\advance\@tempcntb\m@ne\@citeo%
      \@tempcnta\@tempcntc\@tempcntb\@tempcntc\fi\fi}}\@citeo}{#1}}%

\def\@citeo{\ifnum\@tempcnta>\@tempcntb\else\@citea
  \def\@citea{,\penalty\@m}%
  \ifnum\@tempcnta=\@tempcntb\the\@tempcnta\else
   {\advance\@tempcnta\@ne\ifnum\@tempcnta=\@tempcntb \else
\def\@citea{--}\fi
    \advance\@tempcnta\m@ne\the\@tempcnta\@citea\the\@tempcntb}\fi\fi}

\catcode`\@=12

\begin{document}

\title{\bf A Phenomenological Model of the Glasma and Photon Production\footnote{Invited talk presented at the 54'th Cracow School of Theoretical Physics, Zakopane, Poland, June 2014}   }
\author{Larry McLerran$^{(1,2,3)}$ }

\maketitle

\begin{enumerate}

\item Physics Department, 
Brookhaven National Laboratory,
   Upton, NY 11973, USA
 \item RIKEN BNL Research Center, 
 Brookhaven National Laboratory,
   Upton, NY 11973, USA
   
   \item Physics Dept. Central China Normal University, Wuhan, China
\end{enumerate}

\begin{abstract}
I discuss a phenomenological model for the Glasma.  I introduce over occupied distributions for gluons,
and compute their time evolution.
I use this model to estimate the ratio of quarks to gluons  and the entropy production as functions of time.  I then discuss photon production at RHIC and LHC, and how geometric scaling and the Glasma might explain generic features of such production.

 \end{abstract}

\section{Introduction}

There have been many talks at this meeting concerning the Color Glass Condensate\cite{McLerran:1993ka}-\cite{Ferreiro:2001qy} and the Glasma\cite{Kovner:1995ja}-\cite{Lappi:2006fp}, so I will not  present an extended review the subject in this talk.  I will concentrate here on providing a simplified description of  the evolution of the Glasma.  The Glasma is a strongly interacting Quark Gluon Plasma.  It is not thermalized.  It is produced very shortly after the collision of two nuclei, thought of as sheets of Color Glass Condensate,
and evolves into the Thermalized Quark Gluon Plasma.  The Glasma is strongly interacting because the
gluon distributiuons are over occupied, and this overoccupation enhances the interaction strength due to Bose coherence. There may or may not be a Bose condensate of gluons in the Glasma, but this interesting feature will not be the subject of this talk\cite{Blaizot:2011xf}-\cite{Xu:2014ega}.  In fact, I will ignore the possibility of such condensation when I analyze the Glasma, although the result I present may be generalized to the case where condensation is present.

\section{The Glasma}

The Glasma is compoased of gluons that are highly coherent.  The maximal coherence occurs at some momentum scale $\Lambda_{IR}(t)$ where the gluon distributions have strength of order $1/\alpha_s$.
At this infrared scale, the interactions of gluons are maximally strong since the $1/\alpha_S$ in the gluon distributions eats factors of $\alpha_s$ due to gluon interactions.  This is easily seen, since if
the gluon field has strength $1/g$, the coupling strength scales out of classical equations for the gluon fields.

There is also an ultraviolet scale $\Lambda_{UV}(t)$ at which the gluon distribution rapidly goes to zero.
In the early stages of the evolution of the Glasma, the gluons are characterized by only one scale,
and the distribution functions are maximally strong.  We have the initial conditions
\begin{equation}
   \Lambda_{IR}(t_{in}) \sim \Lambda_{UV}(t_{in}) \sim Q_{sat}
    \label{init}
\end{equation}
where $Q_{sat}$ is the typical momentum scale associated with the field in the Color Glass Condensate,
which determines the initial conditions for the Glasma.

As time evolves, both
the infrared and ultraviolet scales change.  Thermalization can oocur when
\begin{equation}
     \Lambda_{IR}(t_{therm}) \sim \alpha_s\Lambda_{UV}(t_{therm}) \sim \alpha_s T_{init}
     \label{therm}
\end{equation}
where $T_{init}$ is the temperature when the system first thermalizes.  One can see this from
thermal Bose distribution functions
\begin{equation}
        f_{BE} = {1 \over {e^{E/T} -1}}
\end{equation}
which for energy much less than the temperature is $f \sim T/E$.  At the UV scale of Eqn. \ref{therm},
the distribution is of order 1, but at the IR, the distributions are of order $1/\alpha_S$  

Usually in thermal field theory it is argued there is a magnetic mass generated at the scale $m_{mag} \sim \alpha_S T$,
which guarantees the distributions do not get too large at small momentum scales.
It might however happen that for the highly occupied distributions typical of heavy ion collisions, that we might genenate a chemical potential for the gluons.  If there are too many gluons, this chemical potential would approach the gluon mass, and the gluon distribution would be infinite at zero momentum.  If one integrates over the gluon distribution, the singularity is integrable, and the number of gluons is a fixed number.  If the number of gluons in our system exceeds this number, then the remainder must go into a Bose condensate.  Whether or not such condensation occurs is a matter of much discussion and controversy, and depends upon dynamical details that are not yet understood\cite{Blaizot:2011xf}-\cite{Xu:2014ega}. We will make the conservative assumption here that no such condensation occurs, although our considerations may be generalized to the case with condensation.

In addition to the fascinating issue of Bose condensation, there are a number of questions that should be asked about the Glasma that we do not yet have firm answers:
\begin{itemize}
\item{How long does it take to thermalize?}
\item{For a  three dimensional Glasma expanding in 1 dimension, how do the longitudinal and transverse pressures depend upon time\cite{Martinez:2010sc}-\cite{Martinez:2010sd}?  This is the system of relevance for heavy ion collisions. How does such a system approach isotropization? }
\item{Such systems have strong fluctuating electric and magnetic fields.  Are there interesting non-perturbative phenomena generated in this weakly coupled system\cite{Gasenzer:2013era}? For example, the strongly coupled
fluctuating fields in the vacuum generate confinement.  Might there be related effects for the chaotic Glasma fields?}
 \end{itemize}
 
 \section{Evolution of the Glasma}
 
There have been a number of attempts to simulate properties of the Glasma.  Early simulations
assumed that the Glasma was uniform in longitudinal coordinate\cite{Krasnitz:1999wc}-\cite{Lappi:2003bi}.  It was soon discovered that such uniformity was destroyed by small fluctuations which led to developing a  turbulent fluid\cite{Mrowczynski:1993qm}-\cite{Romatschke:2005pm}.  The issue then became
how to properly include the quantum fluctuations in the initial conditions which lead to the development of such turbulence, and how fields with  these initial conditions evolve in time.  At present there is consensus on how to set up such a computation\cite{Gelis:2013rba}-\cite{Epelbaum:2014xea}, but not broad consensus on the results of simulations
of the evolution of these fields\cite{Berges:2014yta}.  Classical field methods have difficulties at largish times, and the methods of transport theory have difficulty incuding inelastic effects and properly including condensation phenomena\cite{Epelbaum:2014yja}.  

I think that the results show the promise that although the Glasma may take some time to thermalize,
it may undergo hydrodynamic behaviour from early times.  If so, this hydrodynamics will have a significant anisotropy between longitudinal and transverse pressure\cite{Martinez:2010sc}-\cite{Martinez:2010sd}.  This behaviour is not seen just in Glasma simulations but also in computations employing AdSCFT methods with intrinsic strong coupling\cite{Janik:2005zt}-\cite{Heller:2011ju}.

In what follows, I will construct a simplifed model of the Glasma that illustrates some simple features of the Glasma, and may be useful for phenomenological applications\cite{McLerran:2014hza}.  I will assume that distributions are approximately isotropic, and again the considerations presented here might be generalized to the anisotropic case.

Let us begin with the definition of the gluon distribution function
\begin{equation}
{1 \over {\tau \pi R^2}} {{dN} \over {d^3p}} = f(p)
\end{equation}
where $R$ is the transverse size of the system, and $\tau$ is the proper time.  For a non expanding
system the proper time is just the time, but for a longitudinally expanding system $\tau = \sqrt{t^2-z^2}$.
We take as initial conditions
\begin{equation}
  f(p) \sim {1 \over {\alpha_S}},~ p \le Q_{sat}
\end{equation}
and
\begin{equation}
  f(p) \rightarrow 0, ~p \ge Q_{sat}
\end{equation}
At some point the distribution function must go to zero and will have a value of order 1, so we see that
the UV scale is defined from
\begin{equation}
  f(\Lambda_{UV}) \sim 1
\end{equation}

Generically, the transport eqautions for a highly occupied Bose gas, with $f >> 1$ is of the form
\begin{equation}
  {{df} \over {dt} } \sim \alpha_S^2 f^3
\end{equation}
Implicit in this relationship are integrations on the right hand side of the equation with weight
associated with the scattering kernal.
The factor of $\alpha_S^2$ is the coupling strength.  In scattering there are two particles in the initial
and two particles in the final state, so we would naively expect that the scattering term in the transport equations to be of order $f^4$, but this leading term cancels in the forward and backward going processes
leaving a term of order $f^3$.  

Let us assume that the distribution function is classical for $E << \Lambda_{UV}$, then
\begin{equation}
       f \sim {1 \over \alpha_S} {\Lambda_{IR} \over E}
\end{equation}
More generally we can write
\begin{equation}
  f \sim {1 \over {\alpha_S}} {\Lambda_{IR} \over \Lambda_{UV}} f(E/ \Lambda_{UV})
\end{equation}

Now plugging this into the transport equation and integrating over momentum gives an equation
\begin{equation}
  {d \over dt} \Lambda_{IR} \Lambda_{UV}^2 \sim \Lambda_{IR}^3 \Lambda_{UV}
\end{equation}
Taking
\begin{equation}
  1/t \sim {1 \over {\Lambda_{IR} \Lambda_{UV}^2}} {d \over dt} \Lambda_{IR} \Lambda_{UV}^2
\end{equation}
we can identify the scattering time as
\begin{equation}
  t_{scat} \sim {\Lambda_{UV} \over \Lambda_{IR}^2}
\end{equation}
Note that the coupling constant has entirely disappeared from this equation.  One can show that this
form of the  time dependence persists when one includes higher order corrections associated
 with inelastic particle production\cite{Blaizot:2011xf}.

If there is a Bose condensate present then there is a term in the transport equation associated with scattering from a condensate.  In this case, the dependence upon the infrared and ultraviolet scales for the scattering time is different, but can also be explicitly obtained. 

The relationship between the dynamical scale and the scattering time,  $t \sim t_{scat}$ gives one equation determining the evolution of the scales. The other equation is energy conservation.   The energy density is
\begin{equation}
  \epsilon \sim {1 \over \alpha_S} \Lambda_{IR} \Lambda_{UV}^3
\end{equation}
The solution to these equations in a fixed box or an expanding box gives power law dependences in time for the infrared and ultraviolet scale.

\section{A Simple Model for the Glasma}

It is useful to consider a simple model for the Glasma that is explicit and has the properties 
described above.  Let us take the the gluon distribution function to be an overoccupied Bose-Einstein
distribution\cite{McLerran:2014hza},
\begin{equation}
   f{p) ={\gamma(t) \over {e^{E/\Lambda(t)} -1}}}
\end{equation}
In this form, we see that $\Lambda$ is an effective temperature, and that $\Lambda = \Lambda_{UV}$
The factor $\gamma$ is the overoccupation factor for the Bose-Einstein distribution.  For
a thermally equilibrated distribution $\gamma = 1$.  For the Glasma, we take
\begin{equation}
  \gamma ={1 \over {\alpha_S}}  {\Lambda_{IR} \over \Lambda_{UV}}
\end{equation}
At some time in the evolution
\begin{equation}
 \gamma(t) = 1
 \end{equation}
 At this time, the system is thermal, and the criterion of Eqn. \ref{therm} is satisfied.  At this time, $t_{th}$
is determined from
  \begin{equation}
 T = \Lambda_{UV} (t_{th})
\end{equation}
Beyond this time, $\gamma(t) = 1$, but the temperature may evolve.

The entropy density of these overoccupied distributions is
\begin{equation}
 s = \int~d^3p~ \{ (1+f)ln(1+f) - fln(f) \} \sim \Lambda_{UV}^3 ln\left\{ {\Lambda_{IR} \over {\alpha_S \Lambda_{UV}}} \right\}
\end{equation}
On the other hand the number density of gluons is
\begin{equation}
  \rho \sim {1 \over \alpha_S} \Lambda_{IR} \Lambda_{UV}^2
\end{equation}

The entropy per particle becomes
\begin{equation}
 s/n \sim \alpha_s~ \Lambda_{UV}/\Lambda_{IR} 
\end{equation}
This means that early on when the system is highly coherent, the entropy per particle is small.
By the time of thermalization, the entropy per particle has become of order 1.

We can also estimate the quark to gluon number density.  We take for the quark distribution function
\begin{equation}
  f_{quark} = {1 \over {e^{E/\Lambda(t)} + 1}}
\end{equation}
The quarks cannot be over occupied because they are fermions.  We assume the UV scale is the same for quarks and gluons.  The total number of quarks is
of order
\begin{equation}
  q \sim \Lambda_{UV}^3
 \end{equation} 
 This means that the ratio of quarks to gluons is
 \begin{equation}
   q/g \sim \alpha_S ~\Lambda_{UV}/\Lambda_{IR}
 \end{equation}
 and like the entropy to gluon ratio, it begins small but at thermalization has achieved  a ratio of order
 one.
 This underabundance of quarks at early times has no relatiionship to the rate of quark production.
 It simply reflects the overabundance of gluons, and that Fermi statistics forbid the overoccupation of fermions.
 
 \section{Saturation, the Glasma, and Photons} \begin{figure}
\unitlength\textwidth
\includegraphics[width=1.0\linewidth]{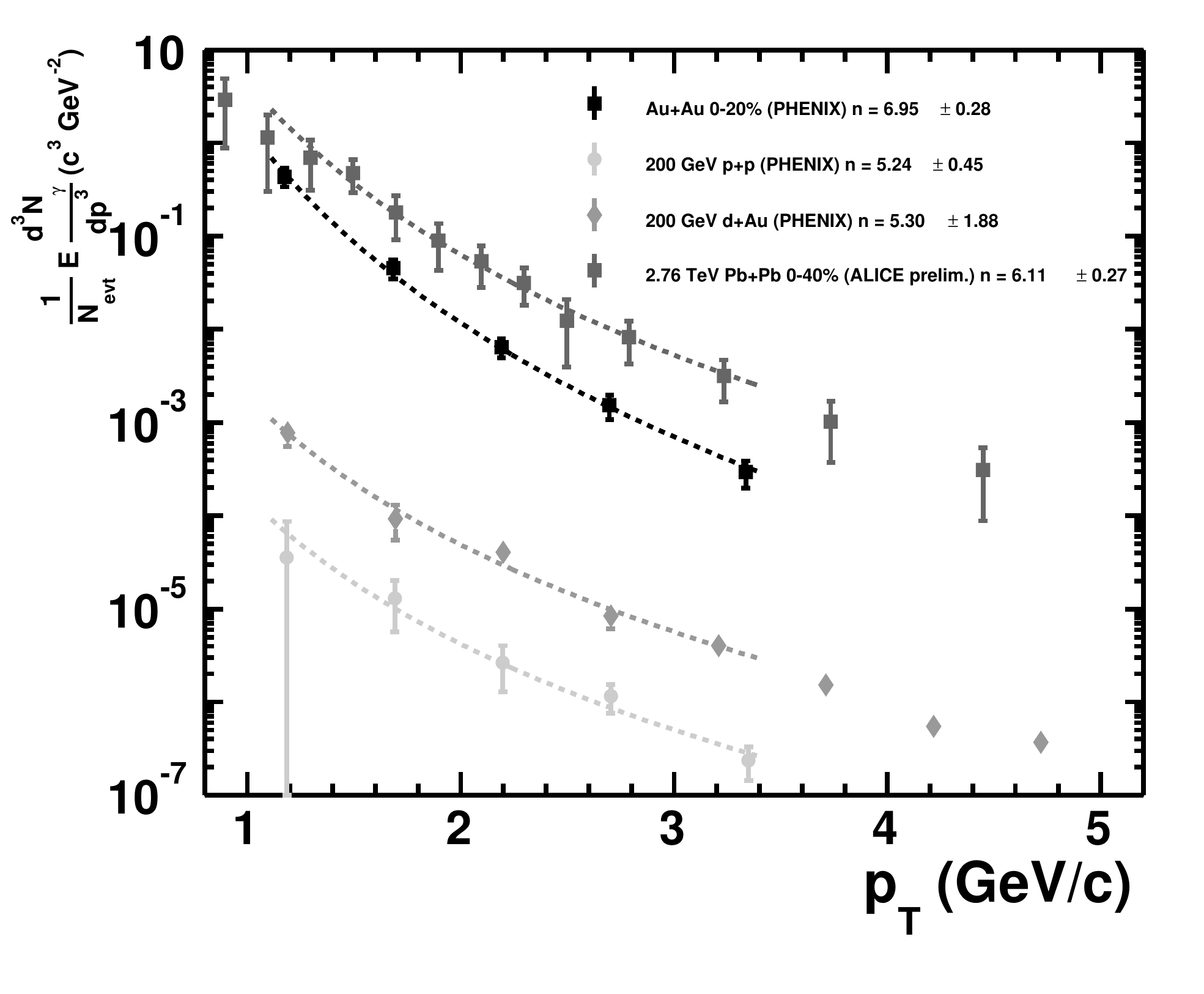}
\caption{Measurements of invariant yields of direct photon production in nuclear collisions below $p_T = 5$~GeV/$c$ compared to power law parameterizations. Data are taken from the PHENIX experiment at RHIC \cite{Adare:2008ab,Adare:2012vn} and the ALICE experiment at the LHC \cite{Wilde:2012wc}. The error bars represent the combined systematic and statistical uncertainties of the measurements.  Original figure is from Ref. \cite{Klein-Bosing:2014uaa}.
\label{fig:spectra} 
}
\end{figure}

\begin{figure}
\unitlength\textwidth
\includegraphics[width=1.0\linewidth]{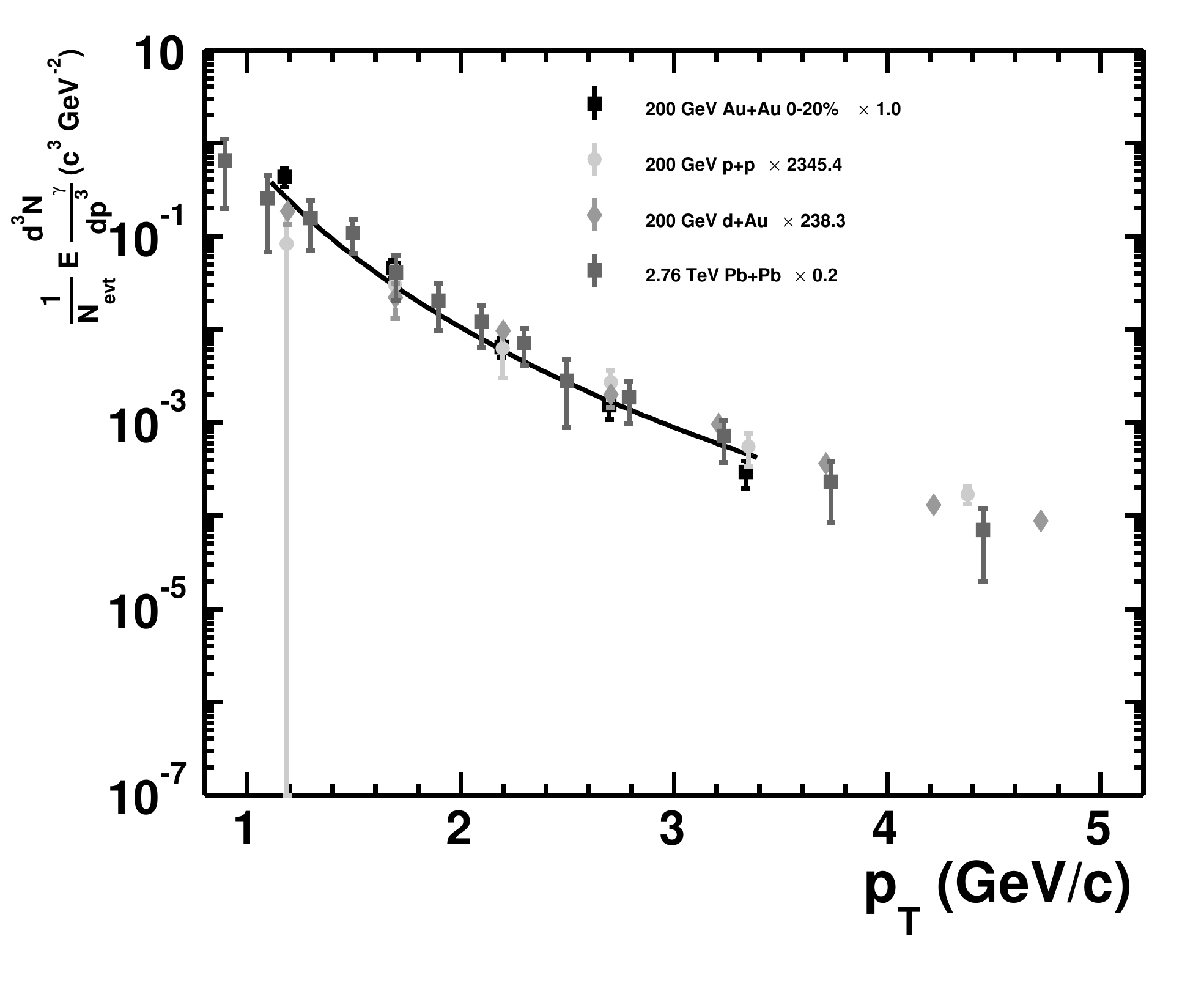}
\caption{Geometrically scaled invariant yields of direct photon production below $p_T = 5$~GeV/$c$, the assumed common power law shape of $p_T^{-6.1}$ has been fit to the PHENIX $AuAu$ data. The error bars represent the combined systematic and statistical uncertainties of the measurements.  Original figure is from Ref. \cite{Klein-Bosing:2014uaa}
\label{fig:spectraScaled} 
}
\end{figure}
 
 If both the Glasma and the Thermalized Quark Gluon Plasma obey approximate hydrodynamic behaviour, it will be difficult to disentangle which is the source of bulk properties of matter produced in heavy ion collisions.  As suggested by Shuryak many years ago\cite{Shuryak:1978ij}, the internal dynamics of an evolving QGP
 might be best addressed by looking at penetrating probes such as photons and dileptons.  These 
 particles can probe the internal dynamics of the QGP and in principle resolve the difference between
 a Glasma and a Thermalized QGP.  It is not easy however, as most experimental observables have
 siginificant contributions from other sources, such as the matter produced at late times as a hadron gas, and from the fragmentation of produced jets into photons.
 
 Nevertheless, we can first try to see if saturation dynamics has anything to do with photon production.
 We can first see whether or not the available photon data has geometric scaling\cite{Stasto:2000er}-\cite{McLerran:2010ex}.  This should be a generic feature of emission from the Color Glass Condensate and early time emission from the Glasma.  In these cases, the only scale in the problem is the saturation momentum  We therefore expect that the distribution of photons will be of the form\cite{Klein-Bosing:2014uaa}
 \begin{equation}
 {1 \over {\pi R^2}} {{d^2N}  \over {dyd^2p_T}} = F \left( {Q_{sat} \over p_T}  \right)
 \end{equation}
 The saturation momenum for nucleus-nucleus collisions is determined by
 \begin{equation}
   Q_{sat}^2 = N_{part}^{1/3} \left( E \over p_T \right)^\delta
 \end{equation}
 Here $N_{part}$ is the number of nucleon participants and $\delta \sim .22-.28$ is determined by
 both fits to deep inelastic scattering data and high energy $pp$ interactions.
 
 The photon data are RHIC and LHC energies is shown in Fig. 1\cite{Klein-Bosing:2014uaa}.  Included are $pp$, 
 $dAu$ and $AuAu$ data from RHIC \cite{Adare:2008ab}-\cite{Adare:2012vn} and $PbPb$ data from LHC\cite{Wilde:2012wc}.  Note that the range of variation of the photon rate is over 4 orders of magnitude.

When we rescale the data using geometric scaling, we obtain the remarkable results of Fig. 2.
It is also true that data from RHIC for $AuAu$ collisions for varying multiplicity of produced particles also falls on this scaling curve.  

The underlying mechanism behind this remarkable scaling behaviour might be jet production and fragmentation into photons\cite{Holopainen:2011pd}-\cite{Qin:2009bk}.  Such a fragmentation process should be approximately scale invariant,
and would preserve the geometric scaling of the initial conditions in the Color Glass Condensate.

We can also try to describe photon production using the Glasma.  Schenke and I used the known lowest order formula
for photon production\cite{Kapusta:1991qp}, with the distribution functions replaced by the over-occupied distribution functions above\cite{McLerran:2014hza}. 
The result is that one can obtain a good description of the spectrum of produced photons in the 1-4 GeV
transverse momentum range.  To do this requires a factor fo 5-10 increase in the rates relative to the
computed rates. Similar results with related meachanisms are found in the semi-QGP analysis of Ref. \cite{Gale:2014dfa}. The Thermalized QGP computations with realistic hydrodynamic simulation are off by a factor of 2-5, so this is a common problem for both computations.

The remarkable result of the photon measurements at RHIC and LHC is the observation that photons flow almost like hadrons.  This is difficult to achieve in Thermalized QGP computations of photon production.  This is because the photons are produced early before much flow develops.  It might be that such photons are produced late in the collision\cite{vanHees:2011vb}-\cite{Linnyk:2013hta}, but then it would be difficult to explain the geometric
 scaling seen in the data.  At very late times there are scales of order $\Lambda_{QCD}$ which become important.  The Glasma is producing significant entropy per gluon during its expansion, and therefore
 cools more slowly than does a Thermalized QGP.  This allows more time for  flow to develop.  It  is possible to get acceptable flow from the Glasma emission, at the expense as mentioned above, of reducing rates of photon emission which are already somewhat low.


\section*{Acknowledgements}
 I thank Michal Praszalowicz for organizing this wonderful meeting.  I also thank the Theoretical Physics Institute at the University of Heidleberg where L. McLerran is a Hans Jensen Professor of physics,
 and where this talk was written up.
 The research L. McLerran  is supported under DOE Contract No. DE-AC02-98CH10886.


\begin{thebibliography} {00}
 
\bibitem{McLerran:1993ka}
  L.~D.~McLerran and R.~Venugopalan,
  Phys.\ Rev.\ D {\bf 49} (1994) 3352
  [hep-ph/9311205].
  
\bibitem{McLerran:1993ni}
  L.~D.~McLerran and R.~Venugopalan,
  Phys.\ Rev.\ D {\bf 49} (1994) 2233
  [hep-ph/9309289].
  
\bibitem{JalilianMarian:1997gr}
  J.~Jalilian-Marian, A.~Kovner, A.~Leonidov and H.~Weigert,
  Phys.\ Rev.\ D {\bf 59} (1998) 014014
  [hep-ph/9706377].
  
\bibitem{Iancu:2000hn}
  E.~Iancu, A.~Leonidov and L.~D.~McLerran,
  Nucl.\ Phys.\ A {\bf 692} (2001) 583
  [hep-ph/0011241].

\bibitem{Ferreiro:2001qy}
  E.~Ferreiro, E.~Iancu, A.~Leonidov and L.~McLerran,
  Nucl.\ Phys.\ A {\bf 703} (2002) 489
  [hep-ph/0109115].


\bibitem{Kovner:1995ja}
  A.~Kovner, L.~D.~McLerran and H.~Weigert,
  Phys.\ Rev.\ D {\bf 52} (1995) 6231
  [hep-ph/9502289].
  
\bibitem{Kovner:1995ts}
  A.~Kovner, L.~D.~McLerran and H.~Weigert,
  Phys.\ Rev.\ D {\bf 52} (1995) 3809
  [hep-ph/9505320].

\bibitem{Kovner:1995ts}
  A.~Kovner, L.~D.~McLerran and H.~Weigert,
  Phys.\ Rev.\ D {\bf 52} (1995) 3809
  [hep-ph/9505320].
  
  
  
  
\bibitem{Krasnitz:1999wc} 
  A.~Krasnitz and R.~Venugopalan,
  Phys.\ Rev.\ Lett.\  {\bf 84}, 4309 (2000)
  [hep-ph/9909203].
  
\bibitem{Krasnitz:2000gz} 
  A.~Krasnitz and R.~Venugopalan,
  Phys.\ Rev.\ Lett.\  {\bf 86}, 1717 (2001)
  [hep-ph/0007108].
  
\bibitem{Krasnitz:2001qu} 
  A.~Krasnitz, Y.~Nara and R.~Venugopalan,
  Phys.\ Rev.\ Lett.\  {\bf 87}, 192302 (2001)
  [hep-ph/0108092].
  
\bibitem{Lappi:2003bi}
  T.~Lappi,
  Phys.\ Rev.\ C {\bf 67} (2003) 054903
  [hep-ph/0303076].
  
  
\bibitem{Lappi:2006fp}
  T.~Lappi and L.~McLerran,
  Nucl.\ Phys.\ A {\bf 772} (2006) 200
  [hep-ph/0602189].
  
\bibitem{Blaizot:2011xf}
  J.~P.~Blaizot, F.~Gelis, J.~F.~Liao, L.~McLerran and R.~Venugopalan,
  Nucl.\ Phys.\ A {\bf 873} (2012) 68
  [arXiv:1107.5296 [hep-ph]].
  
\bibitem{Kurkela:2011ti}
  A.~Kurkela and G.~D.~Moore,
  JHEP {\bf 1112} (2011) 044
  [arXiv:1107.5050 [hep-ph]].
  
\bibitem{Dusling:2010rm}
  K.~Dusling, T.~Epelbaum, F.~Gelis and R.~Venugopalan,
  Nucl.\ Phys.\ A {\bf 850} (2011) 69
  [arXiv:1009.4363 [hep-ph]].
  
\bibitem{Gelis:2013rba}
  T.~Epelbaum and F.~Gelis,
  Phys.\ Rev.\ Lett.\  {\bf 111} (2013) 232301
  [arXiv:1307.2214 [hep-ph]].
  
\bibitem{Epelbaum:2014xea}
  T.~Epelbaum and F.~Gelis,
  Nucl.\ Phys.\ A {\bf 926} (2014) 122
  [arXiv:1401.1666 [hep-ph]].
  
\bibitem{Berges:2013eia}
  J.~Berges, K.~Boguslavski, S.~Schlichting and R.~Venugopalan,
  Phys.\ Rev.\ D {\bf 89} (2014) 074011
  [arXiv:1303.5650 [hep-ph]].
  
\bibitem{Blaizot:2013lga} 
  J.~P.~Blaizot, J.~Liao and L.~McLerran,
  Nucl.\ Phys.\ A {\bf 920}, 58 (2013)
  [arXiv:1305.2119 [hep-ph]].


\bibitem{Huang:2013lia}
  X.~G.~Huang and J.~Liao,
  arXiv:1303.7214 [nucl-th].
  
\bibitem{Xu:2014ega}
  Z.~Xu, K.~Zhou, P.~Zhuang and C.~Greiner,
  arXiv:1410.5616 [hep-ph].
  
\bibitem{Martinez:2010sc} 
  M.~Martinez and M.~Strickland,
  Nucl.\ Phys.\ A {\bf 848}, 183 (2010)
  [arXiv:1007.0889 [nucl-th]].

\bibitem{Martinez:2010sd} 
  M.~Martinez and M.~Strickland,
  Nucl.\ Phys.\ A {\bf 856}, 68 (2011)
  [arXiv:1011.3056 [nucl-th]].
   

\bibitem{Gasenzer:2013era} 
  T.~Gasenzer, L.~McLerran, J.~M.~Pawlowski and D.~Sexty,
  Nucl.\ Phys.\ A {\bf } (2014)
  [arXiv:1307.5301 [hep-ph]].
  
\bibitem{Mrowczynski:1993qm} 
  S.~Mrowczynski,
  Phys.\ Lett.\ B {\bf 314}, 118 (1993).
  
\bibitem{Romatschke:2005pm} 
  P.~Romatschke and R.~Venugopalan,
  Phys.\ Rev.\ Lett.\  {\bf 96}, 062302 (2006)
  [hep-ph/0510121].
  
  
\bibitem{Berges:2014yta} 
  J.~Berges, B.~Schenke, S.~Schlichting and R.~Venugopalan,
  arXiv:1409.1638 [hep-ph].

\bibitem{Epelbaum:2014yja} 
  T.~Epelbaum, F.~Gelis and B.~Wu,
  Phys.\ Rev.\ D {\bf 90}, 065029 (2014)
  [arXiv:1402.0115 [hep-ph]].
  
\bibitem{Janik:2005zt} 
  R.~A.~Janik and R.~B.~Peschanski,
  Phys.\ Rev.\ D {\bf 73}, 045013 (2006)
  [hep-th/0512162].
  
\bibitem{Janik:2006gp} 
  R.~A.~Janik and R.~B.~Peschanski,
  Phys.\ Rev.\ D {\bf 74}, 046007 (2006)
  [hep-th/0606149].
  
\bibitem{Heller:2011ju} 
  M.~P.~Heller, R.~A.~Janik and P.~Witaszczyk,
  Phys.\ Rev.\ Lett.\  {\bf 108}, 201602 (2012)
  [arXiv:1103.3452 [hep-th]].
  
\bibitem{McLerran:2014hza} 
  L.~McLerran and B.~Schenke,
  arXiv:1403.7462 [hep-ph].
  
\bibitem{Shuryak:1978ij} 
  E.~V.~Shuryak,
  Phys.\ Lett.\ B {\bf 78}, 150 (1978)
  [Sov.\ J.\ Nucl.\ Phys.\  {\bf 28}, 408 (1978)]
  [Yad.\ Fiz.\  {\bf 28}, 796 (1978)].
  
\bibitem{Adare:2008ab} 
  A.~Adare {\it et al.}  [PHENIX Collaboration],
  Phys.\ Rev.\ Lett.\  {\bf 104}, 132301 (2010)
  [arXiv:0804.4168 [nucl-ex]].
  
\bibitem{Adare:2012vn} 
  A.~Adare, S.~S.~Adler, S.~Afanasiev, C.~Aidala, N.~N.~Ajitanand, Y.~Akiba, H.~Al-Bataineh and A.~Al-Jamel {\it et al.},
  Phys.\ Rev.\ C {\bf 87}, 054907 (2013)
  [arXiv:1208.1234 [nucl-ex]].
  
\bibitem{Wilde:2012wc} 
  M.~Wilde [ALICE Collaboration],
  Nucl.\ Phys.\ A {\bf 904-905}, 573c (2013)
  [arXiv:1210.5958 [hep-ex]].
  
  
  
\bibitem{Stasto:2000er} 
  A.~M.~Stasto, K.~J.~Golec-Biernat and J.~Kwiecinski,
  Phys.\ Rev.\ Lett.\  {\bf 86}, 596 (2001)
  [hep-ph/0007192].
  
\bibitem{McLerran:2010ex} 
  L.~McLerran and M.~Praszalowicz,
  Acta Phys.\ Polon.\ B {\bf 41}, 1917 (2010)
  [arXiv:1006.4293 [hep-ph]].
  
\bibitem{Klein-Bosing:2014uaa} 
  C.~Klein-Bösing and L.~McLerran,
  Phys.\ Lett.\ B {\bf 734}, 282 (2014)
  [arXiv:1403.1174 [nucl-th]].
  
\bibitem{Holopainen:2011pd} 
  H.~Holopainen, S.~Rasanen and K.~J.~Eskola,
  Phys.\ Rev.\ C {\bf 84}, 064903 (2011)
  [arXiv:1104.5371 [hep-ph]].

\bibitem{Chatterjee:2013naa} 
  R.~Chatterjee, H.~Holopainen, I.~Helenius, T.~Renk and K.~J.~Eskola,
  Phys.\ Rev.\ C {\bf 88}, 034901 (2013)
  [arXiv:1305.6443 [hep-ph]].
  
\bibitem{Qin:2009bk} 
  G.~Y.~Qin, J.~Ruppert, C.~Gale, S.~Jeon and G.~D.~Moore,
  Phys.\ Rev.\ C {\bf 80}, 054909 (2009)
  [arXiv:0906.3280 [hep-ph]].
  
\bibitem{Kapusta:1991qp} 
  J.~I.~Kapusta, P.~Lichard and D.~Seibert,
  Phys.\ Rev.\ D {\bf 44}, 2774 (1991)
  [Erratum-ibid.\ D {\bf 47}, 4171 (1993)].
  
\bibitem{Gale:2014dfa} 
  C.~Gale, Y.~Hidaka, S.~Jeon, S.~Lin, J.-F.~Paquet, R.~D.~Pisarski, D.~Satow and V.~V.~Skokov {\it et al.},
  arXiv:1409.4778 [hep-ph].



  
\bibitem{vanHees:2011vb} 
  H.~van Hees, C.~Gale and R.~Rapp,
  Phys.\ Rev.\ C {\bf 84}, 054906 (2011)
  [arXiv:1108.2131 [hep-ph]].
  
\bibitem{Linnyk:2013hta} 
  O.~Linnyk, V.~P.~Konchakovski, W.~Cassing and E.~L.~Bratkovskaya,
  Phys.\ Rev.\ C {\bf 88}, 034904 (2013)
  [arXiv:1304.7030 [nucl-th]].

\end{thebibliography}
\end{document}